\documentclass[11pt,graphicx,amsmath]{article}
\usepackage{amsmath}
\usepackage{graphicx}
\usepackage{bm}
\usepackage{color}
\usepackage{amssymb}
\usepackage{amsfonts}
\usepackage{comment}
\usepackage{cite}
\usepackage{todonotes}
\usepackage{caption}
\usepackage{subcaption}

\def\bl#1\el{\begin{align}#1\end{align}}

\title{  Analytical Equation of   Three-point Correlation Function of Galaxies:
              to   Third Order of Density Perturbation}

\author{\small  Shu-Guang  Wu\thanks{wusg@mail.ustc.edu.cn}, \,
                Yang Zhang\thanks{yzh@ustc.edu.cn}  \\
\small  Department of  Astronomy,  Key Laboratory
	for Researches in Galaxies and Cosmology, \\
\small  School of Astronomy and Space Sciences, \\
\small  University of Science and Technology of China, Hefei, Anhui, 230026, China \\
}

\date{}

\def\be{\begin{equation}}
\def\ee{\end{equation}}
\def\ba{\begin{eqnarray}}
\def\ea{\end{eqnarray}}
\def\nn{\nonumber}
\def\la{\langle}
\def\ra{\rangle}

\allowdisplaybreaks

\sf

\begin{document}

\maketitle

\begin{abstract}\Large

Applying  functional differentiation
to the density field with Newtonian gravity,
we obtain the static, nonlinear equation of
the three-point correlation function $\zeta$ of galaxies,
to the third order density perturbations.
We make the equation closed and
perform renormalization of the mass and the Jeans wavenumber.
Using the boundary condition inferred from observations,
we obtain the third order solution $\zeta(r, u, \theta)$ at fixed $u=2$,
which is positive,
exhibits a  $U$-shape along the angle $\theta$,
and  decreases monotonously along  the radial $r$
up to the range $r \leq  30\, h^{-1}$Mpc in our computation.
The corresponding reduced $Q(r, u, \theta)$ deviates from 1 of the Gaussian case,
has a deeper $U$-shape along $\theta$,
and varies non-monotonously along $r$.
The  third order solution agrees with the SDSS data of galaxies,
quite close to the previous second order solution,
especially at large scales.
This indicates that
the equations of correlation functions
with increasing orders of density perturbation
provide a stable description of the nonlinear galaxy system.

\end{abstract}


\section{Introduction}

\label{Intro}

In study of the distribution of galaxies,
the n-point correlation functions  (nPCF) are important tools
which contain the dynamical and statistical information of
the system of galaxies
\cite{PeeblesGroth1975,GrothPeebles1977,FryPeebles,Peebles1980,Peebles1993,
Fry1983,Fry1984,Fry1994,Bernardeau2002}.
The analytical, closed equations of the 2PCF  $\xi$
(also denoted as $G^{(2)}$)
up to the second order of density perturbation have been derived
for the static case
\cite{Zhang2007,zhang2009nonlinear,ZhangChen2015,ZhangChenWu2019},
as well as  for the evolution case \cite{ZhangLi2021}.
And the associated solutions have simultaneously provided
simple explanations
of several seemingly-unrelated features of the observed correlation of galaxies,
such as the  power law of the correlation
$\xi \simeq  (r_0/r)^{1.7}$  in a  range $r=(0.1\sim  10) h^{-1}$Mpc,
the  correlation amplitude being proportional to the galaxy mass $\xi \propto m$,
the correlation function of clusters having
a similar form to that of galaxies $\xi_{cc} \simeq (10\sim 20)\, \xi_{gg}$
with a higher amplitude,
the scaling behavior of the cluster correlation amplitude,
the 100Mpc-periodic bumps of the observed $\xi_{gg}(r)$
on  very large scales,
the small wiggles in the power spectrum caused by acoustic oscillating waves,
etc.

The statistic of galaxy distribution
is non-Gaussian due to  long-range gravity,
and $G^{(2)}$ is insufficient to reveal the non-Gaussianity.
It is necessary to study  the 3PCF $G^{(3)}(\mathbf{r, r', r''})$
which statistically describes
the excess probability over random of finding three galaxies located at
the three vertices (${\bf r}$, ${\bf r}'$, ${\bf r}''$)
of a given triangle \cite{FryPeebles,Peebles1980,Fry1984}.
There are a few preliminary analytical studies of $G^{(3)}$.
Ref.\cite{Fry1984} did not give the equation of  $G^{(3)}$
and  tried to calculate $G^{(3)}$ to lowest non-vanishing order in density perturbation,
assuming   initial conditions that are Gaussian and have a power-law spectrum.
Similarly, using the BBGKY hierarchy,
Ref.\cite{Inagaki1991} calculated the Fourier transformation
of $G^{(3)}$ perturbatively under  Gaussian initial conditions.
For the system of galaxies, however,
its non-Gaussian distribution function is unknown,
so that generally one is not able to compute  $G^{(3)}$
even if the density perturbation as one realization  is given.
Besides, the initial power spectrum of the system of galaxies
is not of a simple power-law form even at the early epoch
when galaxies are newly formed at some high redshifts.
Ref.\cite{Bharadwaj19941996} adopted the BBGKY hierarchy method
and formally  wrote down an equation of $G^{(3)}$
for a Newtonian gravity fluid without pressure and vorticity.
But the formal equation contains no pressure and source terms
and will not be  able to exhibit oscillation and clustering properties.
Moreover, the formal equation is not closed yet
and involves other unknown functions beside $G^{(3)}$.
This situation is similar to that in Ref.\cite{DaviesPeebles1977}
which gave an equation for $G^{(2)}$ involving other unknown functions.
These unclosed equations are hard to use
for the actual system of galaxies,
since appropriate initial conditions are difficult to specify
for several unknown functions.
In Ref.\cite{WuZhang2021} the static equation of $G^{(3)}$
was studied to the second order of density perturbation,
and the solution describes the overall profile
of the observed 3PCF  \cite{marin2011}.
In this paper, we shall work on the third order density perturbation,
and also give renormalization of the mass $m$ and the Jeans wavenumber,
and compare the solution with observations.

Within a small redshift range,
the expansion effect is small,
and the correlations of galaxies can be well described by the static equation.
As demonstrated for the case of 2PCF \cite{ZhangLi2021},
 the expansion term in the evolution equation
is about two orders smaller than the pressure and gravity terms,
and the 2PCF increases slowly,
$\xi \propto (1+z)^{-0.2}$ for $z= 0.5 \sim 0.0$.

\section{Equation of 3PCF to third order of density perturbation}
\label{sec:deri3PCF}

The equation of the density field with Newtonian gravity
\cite{Zhang2007,zhang2009nonlinear,ZhangChen2015,ZhangChenWu2019,WuZhang2021}
\be \label{psifieldequ}
\nabla^2 \psi-\frac{(\nabla \psi)^2}{\psi}+k_J^2 \psi^2+J \psi^2 =0 ,
\ee
where $\psi(\mathbf{r}) \equiv  \rho(\mathbf{r}) / \rho_0$
is the rescaled mass density field  with $\rho_0$ being the mean mass density,
and  $k_J \equiv (4\pi G \rho_0/c_s^2)^{1/2}$ is the Jeans wavenumber,
$c_s $ is the sound speed,
and $J$ is the external source employed to carry out functional derivatives
conveniently.
The $n$-point correlation function is defined by
$G^{(n)}({\bf r}_1,\cdots,{\bf r}_n)
 =\la \delta \psi({\bf r}_1) \cdots \delta \psi({\bf r}_n)\ra
 =\frac{1}{\alpha^{n-1}}\frac{\delta^{n-1} \la \psi({\bf r}_1) \ra }
{\delta J({\bf r}_2)\cdots\delta J({\bf r}_n)}\vert_{J=0},
$
where $\delta \psi({\bf r}) = \psi({\bf r}) - \la \psi  \ra$
is the fluctuation around the expectation value $\la \psi \ra $,
and $\alpha=c_s^2/4\pi G m$.
(See Refs.\cite{BinneyDowrickFisherNewman1992,Goldenfeld1992,Zustin1996,
Zhang2007,zhang2009nonlinear,ZhangChen2015,ZhangChenWu2019,ZhangLi2021}.)
To derive the  equation of $G^{(3)}({\bf r}, {\bf r}', {\bf r}'')$,
we take the ensemble average of Eq.(\ref{psifieldequ}) in the presence of $J$,
and take the functional derivative of this equation
twice with respect to the source $J$, and set $J=0$.
The second term in Eq.(\ref{psifieldequ}) is expanded as
\bl \label{secondterm}
\la \frac{(\nabla \psi)^2}{\psi} \ra
 = & \frac{(\nabla \la \psi \ra)^2}{\la \psi \ra}
+ \frac{\la (\nabla \delta \psi)^2 \ra}{\la \psi \ra}
-\frac{\nabla \la \psi \ra}{\la \psi \ra^2}  \cdot \la \nabla (\delta \psi)^2 \ra
+\frac{(\nabla \la \psi \ra)^2}{\la \psi \ra^3} \la (\delta \psi)^2 \ra \nonumber \\
&
-\frac{1}{\la \psi \ra^2} \la \delta \psi (\nabla \delta \psi)^2 \ra
+ \frac{2}{3} \frac{1}{\la \psi \ra^3} \nabla \la \psi
   \ra \cdot \la \nabla (\delta \psi)^3 \ra
-\frac{(\nabla \la \psi \ra)^2}{\la \psi \ra^4} \la (\delta \psi)^3 \ra
+ ,,, .
\el
containing $(\delta \psi)^3$,
higher than our previous work \cite{WuZhang2021}.
Calculations yields the following equation
\ba \label{3PCF}
&& \Big( 1+ \frac{1}{\psi_0^2} G^{(2)}(0) \Big)
  \nabla^2  G^{(3)}(\mathbf{r, r', r''})
+\Big( \frac{2}{\psi_0^2}  \nabla G^{(2)}(0)
- \frac{2}{\psi_0^3}  \nabla G^{(3)}(0) \Big) \cdot
   \nabla G^{(3)}(\mathbf{r, r', r''}) \nonumber \\
& &+ \Big( 2 k_J^2 \psi_0
+ \frac{1}{2 \psi_0^2} \nabla^2  G^{(2)}(0)
-\frac{2}{3 \psi_0^3} \nabla^2 G^{(3)}(0)
-\frac{1}{\psi_0^2}  k_J^2  G^{(3)}(0) \Big) G^{(3)}(\mathbf{r, r', r''})
\nonumber \\
& &+\frac{1}{2 \psi_0^2} G^{(2)}(\mathbf{r, r''})
  \nabla ^2 G^{(3)}(\mathbf{r, r, r'})
+\frac{1}{2 \psi_0^2}  G^{(2)}(\mathbf{r, r'})
    \nabla^2 G^{(3)}(\mathbf{r, r, r''}) \nonumber  \\
& &+\frac{2}{\psi_0^2} \nabla G^{(3)}(\mathbf{r, r, r'})
   \cdot \nabla G^{(2)}(\mathbf{r, r''})
+\frac{2}{\psi_0^2}\nabla  G^{(2)}(\mathbf{r, r'})
  \cdot  \nabla G^{(3)}(\mathbf{r, r, r''})  \nonumber \\
& &+\frac{1}{\psi_0^2} G^{(3)}(\mathbf{r, r,r'}) \nabla^2 G^{(2)}(\mathbf{r, r''})
+\frac{1}{\psi_0^2} G^{(3)}(\mathbf{r, r, r''})
   \nabla^2 G^{(2)}(\mathbf{r, r'})  \nonumber \\
& & -\frac{1}{2 \psi_0} \nabla ^2  G^{(4)}(\mathbf{r, r, r', r''})  \nonumber \\
& & -\frac{2}{3 \psi_0^3} G^{(2)}(\mathbf{r, r''})\nabla^2 G^{(4)}(\mathbf{r, r, r, r'})
-\frac{2}{3 \psi_0^3} G^{(2)}(\mathbf{r, r'})
   \nabla^2 G^{(4)}(\mathbf{r, r, r, r''})   \nonumber \\
& &- \frac{2}{\psi_0^3}  \nabla G^{(2)}(\mathbf{r, r''})
  \cdot  \nabla G^{(4)}(\mathbf{r, r, r, r'})
- \frac{2}{\psi_0^3}  \nabla G^{(2)}(\mathbf{r, r'})
  \cdot \nabla G^{(4)}(\mathbf{r, r, r,r''})
\nonumber \\
& &-\frac{1}{\psi_0^2} k_J^2 G^{(2)}(\mathbf{r, r''}) G^{(4)}(\mathbf{r, r, r, r'})
-\frac{1}{\psi_0^2}  k_J^2  G^{(2)}(\mathbf{r, r'}) G^{(4)}(\mathbf{r, r, r, r''})
\nonumber  \\
&&-\frac{2}{\psi_0} \Big( 1 + \frac{3}{\psi_0^2} G^{(2)}(0)
-\frac{3}{\psi_0^3} G^{(3)}(0) \Big) \nabla G^{(2)}(\mathbf{r, r'})
  \cdot \nabla G^{(2)}(\mathbf{r, r''}) \nonumber \\
& &+\Big( 2 k_J^2   -\frac{1 }{\psi_0^3} \nabla^2  G^{(2)}(0)
   +\frac{2}{\psi_0^4} \nabla^2 G^{(3)}(0)
   +\frac{2}{\psi_0^3}  k_J^2  G^{(3)}(0) \Big) G^{(2)}(\mathbf{r, r'})
    G^{(2)}(\mathbf{r, r''})
\nonumber \\
& &- \Big( \frac{4}{\psi_0^3} \nabla G^{(2)}(0) - \frac{6}{\psi_0^4}
   \nabla G^{(3)}(0) \Big) \cdot \Big( \nabla  G^{(2)}(\mathbf{r, r'})
    G^{(2)}(\mathbf{r, r''})
+ \nabla G^{(2)}(\mathbf{r, r''}) G^{(2)}(\mathbf{r, r'}) \Big)
\nonumber \\
& & -\frac{2}{\psi_0^3}   G^{(2)}(0) \Big( G^{(2)}(\mathbf{r, r'})
   \nabla^2 G^{(2)}(\mathbf{r, r''})
+ G^{(2)}(\mathbf{r, r''}) \nabla^2 G^{(2)}(\mathbf{r, r'}) \Big)
\nonumber \\
&=&- \frac{\psi_0}{\alpha } \big[ \big( 2 - \frac{1}{\psi_0^3}  G^{(3)}(0) \big)
 G^{(2)}(\mathbf{r, r'})
+\frac{1}{ \psi_0^2}  G^{(4)}(\mathbf{r, r, r, r'}) \big]
\delta^{(3)}(\mathbf{r-r''}) \nonumber \\
&&- \frac{\psi_0}{\alpha } \big[ \big( 2 - \frac{1}{\psi_0^3}  G^{(3)}(0) \big)
G^{(2)}(\mathbf{r, r''})
+\frac{1}{ \psi_0^2}  G^{(4)}(\mathbf{r, r, r, r''}) \big]
\delta^{(3)}(\mathbf{r-r'}) ,
\ea
where $ G^{(2)}(0) \equiv G^{(2)}(\mathbf{r, r})$,
$G^{(3)}(\mathbf{r, r, r}) \equiv G^{(3)}(0)$,
$\psi_0 \equiv  \la \psi  \ra |_{J=0}=1$,
and $\nabla \equiv \nabla_{\mathbf{r}}$.
We have  neglected $G^{(5)}$ as a cutoff of the hierarchy.
Comparing  with the  2nd-order equation \cite{WuZhang2021},
eq.(\ref{3PCF}) contains also $G^{(2)} G^{(4)}$ terms,
and $ G^{(4)}$ in the delta source.
As expected,   eq.(\ref{3PCF}) reduces to that of
the Gaussian approximation \cite{ZhangChenWu2019},
when all the higher order terms,
such as $ G^{(2)}  G^{(2)}G^{(2)}$,  $ G^{(2)} G^{(3)}$, $G^{(4)}$, are dropped,
Since eq.(\ref{3PCF})  contains $G^{(4)}$, it is  not closed for $G^{(3)}$.
To cutoff the hierarchy,
we adopt Fry-Peebles ansatz \cite{FryPeebles}
\ba \label{frypeeblesansatz}
G^{(4)}(\mathbf{r}_1, \mathbf{r}_2, \mathbf{r}_3, \mathbf{r}_4)
&=&R_a[ G^{(2)}(\mathbf{r}_1, \mathbf{r}_2)
G^{(2)}(\mathbf{r}_2, \mathbf{r}_3) G^{(2)}(\mathbf{r}_3, \mathbf{r}_4)
   +\mathrm{sym. (12 \, \, terms)} ] \nonumber \\
& &+R_b[ G^{(2)}(\mathbf{r}_1, \mathbf{r}_2)
G^{(2)}(\mathbf{r}_1, \mathbf{r}_3) G^{(2)}(\mathbf{r}_1, \mathbf{r}_4)
   +\mathrm{sym. (4 \, \, terms)} ],
\ea
where   $R_a$ and $R_b$ are dimensionless constants,
and $(3R_a+R_b)/4 \simeq 2.5 \pm 0.5$ as constrained by observations
\cite{Fry1983,Fry1984,Szapudi1992, Meiksin1992,Peebles1993}.
The ansatz  \eqref{frypeeblesansatz} leads to
\ba \label{g4rr}
G^{(4)}(\mathbf{r, r, r', r''})
&=&2 R_a G^{(2)}(0) G^{(2)}(\mathbf{r, r'}) G^{(2)}(\mathbf{r', r''})
+2 R_a G^{(2)}(0) G^{(2)}(\mathbf{r, r''}) G^{(2)}(\mathbf{r', r''}) \nonumber \\
& &+2 (R_a + R_b) G^{(2)}(0) G^{(2)}(\mathbf{r, r'}) G^{(2)}(\mathbf{r, r''})
+2 R_a G^{(2)}(\mathbf{r, r'}) G^{(2)}(\mathbf{r, r''})
      G^{(2)}(\mathbf{r', r''})  \nonumber \\
& &+2 R_a  G^{(2)}(\mathbf{r, r'})^2 G^{(2)}(\mathbf{r, r''})
+2 R_a G^{(2)}(\mathbf{r, r'}) G^{(2)}(\mathbf{r, r''})^2   \nonumber \\
& &+R_b G^{(2)}(\mathbf{r, r'})^2 G^{(2)}(\mathbf{r', r''})
+R_b G^{(2)}(\mathbf{r, r''})^2 G^{(2)}(\mathbf{r', r''}) ,
\ea
and
\be \label{g4rrr1}
G^{(4)}(\mathbf{r, r, r, r'})
=3 (2 R_a + R_b )G^{(2)}(0)^2 G^{(2)}(\mathbf{r, r'})
+6 R_a G^{(2)}(0) G^{(2)}(\mathbf{r, r'})^2
+R_b G^{(2)}(\mathbf{r, r'})^3.
\ee
Eq.\eqref{3PCF} also contains  the squeezed
$G^{(3)}(\mathbf{r, r, r'})=\lim\limits_{{\bf r}''
	\rightarrow {\bf r}}G^{(3)}(\mathbf{r, r', r''})$
with  three points being reduced to two.
In observations and simulations $G^{(3)}(\mathbf{r, r, r'})$ can not be resolved,
\cite{Gaztanaga2005,  McBride2011a, McBride2011b,  Yuan2017}.
To avoid the divergence,
we adopt the  Groth-Peebles ansatz \cite{GrothPeebles1977,PeeblesGroth1975}
\be \label{Groth-Peebles-ansatz}
G^{(3)}(\mathbf{r, r', r''})=Q[ G^{(2)}(\mathbf{r, r'}) G^{(2)}(\mathbf{r', r''})
+G^{(2)}(\mathbf{r', r''}) G^{(2)}(\mathbf{r'', r})
+G^{(2)}(\mathbf{r'', r}) G^{(2)}(\mathbf{r, r'})] ,
\ee
where the constant $Q \sim 1$ as constrained by observations.
Then the squeezed 3PCF becomes
\be \label{g3rrr'}
G^{(3)}(\mathbf{r, r, r'})
=2 Q G^{(2)}(0) G^{(2)}(\mathbf{r, r'})+Q G^{(2)}(\mathbf{r, r'})^2,
\ee
consisting of  regular   $ G^{(2)}(\mathbf{r, r'})$.
Substituting  \eqref{g4rr} \eqref{g4rrr1}  \eqref{g3rrr'} into
eq.(\ref{3PCF}),
we obtain the closed equation of the 3PCF
\begin{eqnarray}
	&& \nabla^2  G^{(3)}(\mathbf{r, r', r''})
	+{\bf a}^{(3)}\cdot \nabla G^{(3)}(\mathbf{r, r', r''})
	+  2 g^{(3)}k_J^2 G^{(3)}(\mathbf{r, r', r''})
-\mathcal{A}^{(3)}(\mathbf{r, r', r''})  \nn \\
	&=&-\frac{1}{\alpha} \bigg(2
	- \big( 1+ b \big) e + 3 (2 R_a + R_b )b^2
	+6 R_a bG^{(2)}(\mathbf{r, r''})
	+R_b G^{(2)}(\mathbf{r, r''})^2 \bigg)
	\delta^{(3)}(\mathbf{r-r'}) G^{(2)}(\mathbf{r, r''})\nonumber \\
	&&-\frac{1}{\alpha} \bigg(2
	- \big( 1+ b \big) e  + 3 (2 R_a + R_b )b^2
	+6 R_a b G^{(2)}(\mathbf{r, r'})
	+R_b G^{(2)}(\mathbf{r, r'})^2 \bigg)
	\delta^{(3)}(\mathbf{r-r''}) G^{(2)}(\mathbf{r, r'}), \nonumber \\
 \label{3PCF_02}
\end{eqnarray}
where
\begin{eqnarray}
	&& \mathcal{A}^{(3)}(\mathbf{r, r', r''}) \nn \\
	& = & \bigg[ \bigg( 2 + 2 \big(3
	+ R_a + R_b - 4 Q  \big) b + 12 (2 R_a + R_b ) b^2 \bigg)\big( 1+ b \big)^{-1} 	
	-6 e\bigg] \nabla G^{(2)}(\mathbf{r, r'})
         \cdot \nabla G^{(2)}(\mathbf{r, r''})  \nonumber \\
	& &-\bigg[2 k_J^2 \big( 1+ b \big)^{-1}
	-6 k_J^2 (2 R_a + R_b ) b^2 \big( 1+ b \big)^{-1}
	+ 2  k_J^2  e
	- \bigg( 1  + R_a + R_b- 2 Q + 8  (2 R_a + R_b ) b \bigg) c
	 \nonumber \\
	&& - 2 (2 R_a + R_b ) \big( 1+ b \big) \lvert \mathbf{a}^{(2)} \rvert^2
	+ 2 f  \bigg] G^{(2)}(\mathbf{r, r'})   G^{(2)}(\mathbf{r, r''})
	+  R_a c G^{(2)}(\mathbf{r', r''}) \big( G^{(2)}(\mathbf{r, r'})
	+ G^{(2)}(\mathbf{r, r''}) \big)	 \nonumber \\
	& &+ \bigg[ \bigg( R_a + R_b - 3 Q -1 +  10 (2 R_a + R_b ) b \bigg) \mathbf{a}^{(2)}
	+3 \mathbf{a}^{(3)} \bigg] \cdot \bigg( \nabla G^{(2)}(\mathbf{r, r'})
        G^{(2)}(\mathbf{r, r''})
	+ \nabla G^{(2)}(\mathbf{r, r''}) G^{(2)}(\mathbf{r, r'}) \bigg)  \nonumber \\
	& & +\bigg( 2 - 3 Q + R_a + R_b
	+ 2 (2 R_a + R_b ) b \bigg) \frac{b}{1+ b}
	\bigg( \nabla^2 G^{(2)}(\mathbf{r, r'}) G^{(2)}(\mathbf{r, r''})
	+ \nabla^2 G^{(2)}(\mathbf{r, r''}) G^{(2)}(\mathbf{r, r'}) \bigg) \nonumber \\
	&&+ R_a G^{(2)}(\mathbf{r', r''})
         \bigg[\frac{b}{1+ b} \big( \nabla^2 G^{(2)}(\mathbf{r, r'})
	+ \nabla^2 G^{(2)}(\mathbf{r, r''}) \big)
	+  \mathbf{a}^{(2)} \cdot
	\big( \nabla G^{(2)}(\mathbf{r, r'})
	+ \nabla G^{(2)}(\mathbf{r, r''}) \big)  \bigg] \nonumber \\
	& &+\big( 2 R_a + 8 R_a b - Q \big) \big( 1+ b \big)^{-1}
       \bigg[ \bigg( \lvert \nabla G^{(2)}(\mathbf{r, r'}) \rvert^2
	+ G^{(2)}(\mathbf{r, r'}) \nabla^2 G^{(2)}(\mathbf{r, r'}) \bigg)
       G^{(2)}(\mathbf{r, r''})  \nonumber \\
	&&+  \bigg( \lvert \nabla G^{(2)}(\mathbf{r, r''}) \rvert^2
	+ G^{(2)}(\mathbf{r, r''}) \nabla^2 G^{(2)}(\mathbf{r, r''}) \bigg)
	G^{(2)}(\mathbf{r, r'})  \bigg]  \nonumber \\
	&&+ \big(R_a - Q\big) \big(1+ b\big)^{-1} \bigg[4 \big( G^{(2)}(\mathbf{r, r'})
	+ G^{(2)}(\mathbf{r, r''}) \big) \nabla G^{(2)}(\mathbf{r, r'})
         \cdot   \nabla G^{(2)}(\mathbf{r, r''})  \nonumber \\
	&&+ G^{(2)}(\mathbf{r, r'})^2 \nabla^2 G^{(2)}(\mathbf{r, r''})
	+ \nabla^2 G^{(2)}(\mathbf{r, r'}) G^{(2)}(\mathbf{r, r''})^2 \bigg]
	\nonumber \\
	&& + 24 R_a b \big(1+ b\big)^{-1}  \big( G^{(2)}(\mathbf{r, r'})
	+ G^{(2)}(\mathbf{r, r''}) \big) \nabla G^{(2)}(\mathbf{r, r'}) \cdot
       \nabla G^{(2)}(\mathbf{r, r''})  \nonumber \\
	& &+\bigg(  4 R_a c + 6 k_J^2 R_a b \big( 1+ b \big)^{-1} \bigg)
      \big( G^{(2)}(\mathbf{r, r'})
	+ G^{(2)}(\mathbf{r, r''}) \big)  G^{(2)}(\mathbf{r, r'})
          G^{(2)}(\mathbf{r, r''})  \nonumber \\
	&&+ R_a  \mathbf{a}^{(2)} \cdot \bigg[8 \bigg( \nabla G^{(2)}(\mathbf{r, r'})
       +  \nabla G^{(2)}(\mathbf{r, r''}) \bigg)  G^{(2)}(\mathbf{r, r'})
        G^{(2)}(\mathbf{r, r''}) \nonumber \\
	&&+ 6  \bigg( \nabla G^{(2)}(\mathbf{r, r''})  G^{(2)}(\mathbf{r, r'})^2
	+ \nabla G^{(2)}(\mathbf{r, r'})  G^{(2)}(\mathbf{r, r''})^2 \bigg)\bigg] \nonumber \\
	& &+ R_a \big(1+b\big)^{-1} G^{(2)}(\mathbf{r', r''})
       \bigg( \nabla^2 G^{(2)}(\mathbf{r, r'}) G^{(2)}(\mathbf{r, r''})
	+ G^{(2)}(\mathbf{r, r'}) \nabla^2 G^{(2)}(\mathbf{r, r''})	 \nonumber \\
	&&+ 2 \nabla G^{(2)}(\mathbf{r, r'}) \cdot \nabla G^{(2)}(\mathbf{r, r''}) \bigg)
	+ R_b \big(1+ b\big)^{-1}  G^{(2)}(\mathbf{r', r''})
        \bigg( \lvert \nabla G^{(2)}(\mathbf{r, r'}) \rvert^2 \nonumber \\
	&&+G^{(2)}(\mathbf{r, r'}) \nabla^2 G^{(2)}(\mathbf{r, r'})
	+ \lvert \nabla G^{(2)}(\mathbf{r, r''}) \rvert^2
	+G^{(2)}(\mathbf{r, r''}) \nabla^2 G^{(2)}(\mathbf{r, r''}) \bigg) \nonumber \\
	&&+2 R_b \big(1+ b\big)^{-1}  \bigg[ G^{(2)}(\mathbf{r, r'}) G^{(2)}(\mathbf{r, r''})
	\big(G^{(2)}(\mathbf{r, r'})  \nabla^2 G^{(2)}(\mathbf{r, r'})
	+ G^{(2)}(\mathbf{r, r''})  \nabla^2 G^{(2)}(\mathbf{r, r''})   \nonumber \\
	& & + 2 \lvert \nabla G^{(2)}(\mathbf{r, r'}) \rvert^2
	+2 \lvert \nabla G^{(2)}(\mathbf{r, r''}) \rvert^2 \big)
	+ 3 \big( G^{(2)}(\mathbf{r, r'})^2+G^{(2)}(\mathbf{r, r''})^2 \big)
    \nabla G^{(2)}(\mathbf{r, r'}) \cdot  \nabla  G^{(2)}(\mathbf{r, r''}) \bigg]
            \nonumber \\
	&&+ R_b k_J^2  \big(1+ b\big)^{-1}  \big( G^{(2)}(\mathbf{r, r'})^2
	+ G^{(2)}(\mathbf{r, r''})^2 \big) G^{(2)}(\mathbf{r, r'})
           G^{(2)}(\mathbf{r, r''}) ,
  \label{Adef}
\end{eqnarray}
In eqs.\eqref{3PCF_02} and \eqref{Adef},
${\bf a}^{(3)}\equiv (1+b)^{-1} (\frac{2}{\psi_0^2}
\nabla G^{(2)}(0)- \frac{2}{\psi_0^3}  \nabla G^{(3)}(0) )$,
 $\mathbf{a}^{(2)} \equiv (1+b)^{-1} \frac{2}{\psi_0^2} \nabla G^{(2)}(0)$,
$b \equiv  \frac{1}{\psi_0^2} G^{(2)}(0)$,
$c \equiv \nabla^2 G^{(2)}(0) / [(1+b) \psi_0^2]$,
$e \equiv  G^{(3)}(0) / [(1+b) \psi_0^3]$,
$f \equiv \nabla^2 G^{(3)}(0) / [(1+b) \psi_0^3]$,
$g^{(3)} =\frac{1}{1+b}  + \frac{c}{4 k_J^2 }
-\frac{f}{3 k_J^2} -\frac{e}{2}$,
and  $\alpha$ has absorbed a factor  $(1+b)$.
The six constants,
${\bf a}^{(3)}, {\bf a}^{(2)},  b,c,e,f,$
are combinations of six unknowns:
$G^{(2)}(0)$, $G^{(3)}(0)$,
$\nabla G^{(2)}(0)$, $\nabla G^{(3)}(0)$,
$\nabla^2 G^{(2)}(0)$ and $\nabla^2 G^{(3)}(0)$,
which  can be formally divergent, and are not directly measurable.
These constants are inevitable in the perturbation approach to
any field theory with interactions,
and are often treated by some renormalization.
In our case,
we shall set $g^{(3)} =1$ as the renormalization
of the Jeans wavenumber $k_J$,
and take $(1+b)m$ as the renormalized mass.
Eq.(\ref{3PCF_02}) is a generalized Poisson equation  \cite{Hackbusch2017}
with the two delta sources located at $\bf r'$ and $\bf r''$ respectively,
and the inhomogeneous term $\mathcal{A}^{(3)}$.
Its  structure is similar to
the second order equation \cite{WuZhang2021},
but  $\mathcal{A}^{(3)}$ has more terms.
It also contains
a convection term $\mathbf{a}^{(3)} \cdot \nabla  G^{(3)}(\mathbf{r, r', r''})$
and a gravitating term $g^{(3)} k_J^2 G^{(3)}(\mathbf{r, r', r''})$.
The Jeans wavenumber $k_J$ determines the 3-point correlation length
of the system of galaxies.
 $\alpha^{-1} \propto m$
 determines the  correlation amplitude at small scales,
so that massive galaxies will have a higher amplitude of $G^{(3)}$.
These two properties are  analogous to those of 2PCF \cite{Zhang2007,ZhangLi2021}.

\begin{figure}[htb]
	\centering
	\includegraphics[width = .7\linewidth]{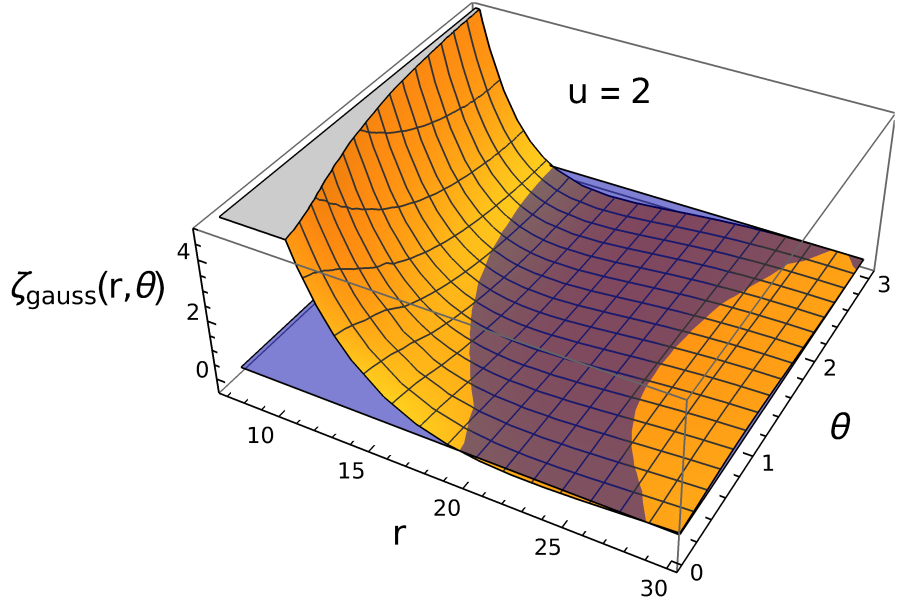}
	\caption{ The Gaussian     $\zeta(r, u, \theta )$
        in the spherical coordinate with $u=2$.
	Along the $r$-direction,
    $\zeta(r)$  becomes negative around $r = (14\sim 27) h^{-1}$Mpc
    forming  a shallower $U$-shape.
    Along the $\theta$-direction,
    $\zeta(\theta )$ decreases monotonously     at small $r$,
    and oscillates at large $r$.
	}
	\label{zetagauss}
\end{figure}

When all nine nonlinear parameters are neglected,
eq.(\ref{3PCF_02}) reduces to the Gaussian approximation
as the next order to the mean field theory \cite{Zhang2007,ZhangChenWu2019},
$G^{(2)}( {\mathbf r}_1, { \mathbf r}_2)
 \propto  \frac{\cos(\sqrt2  k_J \, r_{12})}{ r_{12}}$
with  $r_{12} =|{\mathbf r}_1 - { \mathbf r}_2|$, and
$G^{(3)}(\mathbf{r, r', r''})$
 given by  the Groth-Peebles ansatz \eqref{Groth-Peebles-ansatz} with $Q=1$.
We plot the Gaussian  $G^{(3)}$ in Fig.\ref{zetagauss}.
Here the Gaussian approximation of the self-gravity density field
is conceptually not the same as
the Gaussian random process in statistics.
A ``reduced"  3PCF is often introduced  as the following
\cite{Jing2004, Wang2004, Gaztanaga2005,
Nichol2006, McBride2011a, McBride2011b, Guo2013, Guo2016}
\be \label{grothpeeblesansatz}
Q(\mathbf{r, r', r''})
\equiv \frac{G^{(3)}(\mathbf{r, r', r''})}
{G^{(2)}(\mathbf{r, r'})  G^{(2)}(\mathbf{r', r''})
	+G^{(2)}(\mathbf{r', r''}) G^{(2)}(\mathbf{r'', r})
	+G^{(2)}(\mathbf{r'', r}) G^{(2)}(\mathbf{r, r'})} .
\ee
which is an extension of the Groth-Peebles ansatz \eqref{Groth-Peebles-ansatz}.
$Q(\mathbf{r, r', r''})\ne 1$ is a criterion of non-Gaussianity.

\section{Solution and   comparison with observations }
\label{sec:sol3PCF}

In a homogeneous and isotropic Universe,
 $G^{(2)}(\mathbf{r, r'})= G^{(2)}(\mathbf{|r- r'|})$.
The 3PCF  is  parametrized by
$G^{(3)}(\mathbf{r, r', r''}) \equiv  \zeta(r, u, \theta )$,
where
$ r_{12} \equiv r$,
$u=\frac{r_{1 3}}{r_{12}}$,
$\theta =\cos^{-1}(\hat{{\bf r}}_{12}\cdot\hat{{\bf r}}_{13})$ \cite{marin2011}.
For convenience,
we take ${ \mathbf r}''=\mathbf{0}$
and put the vector $\mathbf{r}'-\mathbf{r}''={ \mathbf r}'$
along the polar axis (see Fig.1 in Ref.\cite{WuZhang2021}),
and write
$G^{(2)}(\mathbf{r, r''}) =\xi(r),$
$G^{(2)}(\mathbf{r, r'}) =\xi(l)$,
$G^{(2)}(\mathbf{r', r''}) =\xi(r')=\xi(u r)$,
where
$l \equiv \lvert \mathbf{r}- \mathbf{r'} \rvert =\beta r$,
$\beta \equiv \sqrt{1+u^2-2u \cos \theta}$.
Then eq. (\ref{3PCF_02}) is written  in spherical coordinate
\bl \label{3PCF_02sp}
&\frac{1}{r^2}  \frac{\partial}{\partial r}
\big(r^2 \frac{\partial}{\partial r}\zeta(r, u, \theta) \big)
+\frac{1}{r^2 \sin \theta}\frac{\partial}{\partial \theta}
\big(\sin \theta \frac{\partial \zeta(r, u, \theta)}{\partial \theta} \big)
\nn \\
& +a^{(3)}_r \frac{\partial \zeta(r, u, \theta)}{\partial r}
  +2  k_J^2 \zeta(r, u, \theta)
- \mathcal{A}^{(3)}(r, u, \theta) \nonumber \\
& = -\frac{b}{\alpha} \Big(2- ( 1+ b ) e
    + 4 (3 R_a + R_b )b^2    \Big)
    \Big(  \frac{2}{\lvert 1-u \rvert } \frac{\delta(\theta)}{\sin \theta} + 1 \Big)
\frac{\delta(r)}{4 \pi r^2},
\el
where  $a^{(3)}_r$ is the $r-$component of $\mathbf{a}^{(3)}$,  and
\begin{eqnarray} \nonumber
	&&\mathcal{A}^{(3)}(r,u,\theta) \nonumber \\
	&=&\bigg[ \bigg( 2 + 2 \big(3
	+ R_a + R_b - 4 Q  \big) b + 12 (2 R_a + R_b ) b^2 \bigg) \big( 1+ b \big)^{-1} 	
	-6 e\bigg] \beta \xi'(l) \xi'(r)  \nonumber \\
	& &-\bigg[2 k_J^2 \big( 1+ b \big)^{-1}
            - \bigg( 1  + R_a + R_b- 2 Q + 8  (2 R_a + R_b ) b \bigg) c
	- 2 (2 R_a + R_b ) \big( 1+ b \big) (a^{(2)}_r)^2 \nonumber \\
	&& -6 k_J^2 (2 R_a + R_b ) b^2 \big( 1+ b \big)^{-1}
          + 2 f + 2  k_J^2  e  \bigg] \xi(l) \xi(r)
	\nonumber \\
	& &+ \bigg[ \bigg( R_a + R_b - 3 Q -1 +  10 (2 R_a + R_b ) b \bigg) a^{(2)}_r
	+3 a^{(3)}_r \bigg] \bigg(\beta \xi'(l) \xi(r) + \xi(l) \xi'(r)\bigg)\nonumber \\
	& & 	+\bigg( 2 - 3 Q + R_a + R_b
	+ 2 (2 R_a + R_b ) b \bigg) \frac{b}{1+ b}
	\bigg[ \big( \frac{2}{r} \xi'(r) + \xi''(r)   \big)\xi(l)  \nonumber \\
	& &	+\bigg( \big( \frac{2}{r} \beta +\frac{2 u}{\beta r}\cos \theta
	-\frac{u^2 \sin^2 \theta }{\beta^3 r} \big) \xi'(l)
	+\big( \beta^2 + \frac{u^2}{\beta^2} \sin^2 \theta \big) \xi''(l) \bigg) \xi(r)
	 \bigg] \nonumber \\
	& &+\big( 2 R_a + 8 R_a b - Q \big) \big( 1+ b \big)^{-1}
         \bigg\{ \big(\beta^2 + \frac{u^2 \sin^2 \theta}{\beta^2}\big)\xi'(l)^2 \xi(r)
	+  \xi(l) \xi'(r)^2 \nonumber \\
	&&+\bigg[ \big( \frac{2}{r} \beta +\frac{2 u}{\beta r}\cos \theta
	-\frac{u^2 \sin^2 \theta }{\beta^3 r} \big) \xi'(l)
	+\big( \beta^2 + \frac{u^2}{\beta^2} \sin^2 \theta \big) \xi''(l)
        + \frac{2}{r} \xi'(r) + \xi''(r)  \bigg]
	\xi(l) \xi(r)   \bigg\} \nonumber \\
	&&+\big( R_a - Q \big) \big( 1+ b \big)^{-1}
         \bigg[ \bigg(\big( \frac{2}{r} \beta +\frac{2 u}{\beta r}\cos \theta
	-\frac{u^2 \sin^2 \theta }{\beta^3 r} \big) \xi'(l) 	
	+\big( \beta^2 + \frac{u^2}{\beta^2} \sin^2 \theta \big)
       \xi''(l)\bigg) \xi(r)^2 \nonumber \\
	&&+\xi(l)^2 \big( \frac{2}{r} \xi'(r) + \xi''(r)  \big)
	+4 \big( \xi(l)  + \xi(r) \big) \beta \xi'(l) \xi'(r) \bigg]
	\nonumber \\
	&& + R_a b \big( 1+ b \big)^{-1} \bigg[ 24\big( \xi(l)
	+ \xi(r) \big) \beta \xi'(l) \xi'(r)
	+\xi(u \, r) \bigg(\big( \frac{2}{r} \beta +\frac{2 u}{\beta r}\cos \theta
	-\frac{u^2 \sin^2 \theta }{\beta^3 r} \big) \xi'(l) \nonumber \\
	&& +\big( \beta^2 + \frac{u^2}{\beta^2} \sin^2 \theta \big) \xi''(l) 	
	+ \frac{2}{r} \xi'(r) + \xi''(r)   \bigg) \bigg]
	+\bigg(  4 R_a c + 6 k_J^2 R_a b / \big( 1+ b \big) \bigg) \big( \xi(l)
	+ \xi(r) \big)  \xi(l) \xi(r) \nonumber \\
	&&+ R_a  a^{(2)}_r \bigg[8 \big( \beta \xi'(l)  + \xi'(r) \big)   \xi(l) \xi(r)
	+ 6 \big( \xi'(r)  \xi(l)^2
	+ \beta \xi'(l)  \xi(r)^2 \big)
	+ \xi(u \, r) \big( \beta \xi'(l)  + \xi'(r) \big)\bigg] \nonumber \\
	&&+ \frac{R_b}{ 1+ b } \bigg\{ \xi(u \, r)
         \bigg[ \big(\beta^2 + \frac{u^2 \sin^2 \theta}{\beta^2}\big)\xi'(l)^2
	+ \xi'(r)^2
	+ \big( \frac{2}{r} \xi'(r) + \xi''(r) \big) \xi(r) \nonumber \\
	& &	+  \bigg( \big( \frac{2}{r} \beta +\frac{2 u}{\beta r}\cos \theta
	-\frac{u^2 \sin^2 \theta }{\beta^3 r} \big) \xi'(l)
	+\big( \beta^2 + \frac{u^2}{\beta^2} \sin^2 \theta \big)
       \xi''(l) \bigg) \xi(l) \bigg]\nonumber \\
	&&+6\big( \xi(l)^2+\xi(r)^2 \big) \beta \xi'(l) \xi'(r)
	+2 \xi(l) \xi(r) \bigg[ 2\bigg( \big(\beta^2
         + \frac{u^2 \sin^2 \theta}{\beta^2}\big)\xi'(l)^2
	+ \xi'(r)^2 \bigg) \nonumber \\
	&& +\bigg( \big( \frac{2}{r} \beta +\frac{2 u}{\beta r}\cos \theta
	-\frac{u^2 \sin^2 \theta }{\beta^3 r} \big) \xi'(l)
	+\big( \beta^2 + \frac{u^2}{\beta^2} \sin^2 \theta \big)
        \xi''(l) \bigg) \xi(l) + \big(\frac{2}{r} \xi'(r) +
        \xi''(r) \big) \xi(r) \bigg] \bigg\}\nonumber  \\
	&&+\frac{R_b k_J^2}{ 1+ b } \big( \xi(l)^2 + \xi(r)^2 \big) \xi(l) \xi(r)
	+  R_a c \xi(u \, r) \big( \xi(l) + \xi(r) \big) 	 \nonumber \\
	& &+\frac{R_a}{ 1+ b } \xi(u \, r)
        \bigg[ \bigg(\big( \frac{2}{r} \beta +\frac{2 u}{\beta r}\cos \theta
	-\frac{u^2 \sin^2 \theta }{\beta^3 r} \big) \xi'(l)
	+\big( \beta^2 + \frac{u^2}{\beta^2} \sin^2 \theta \big)
          \xi''(l)\bigg) \xi(r) \nonumber \\
	&&+ \xi(l) \big( \frac{2}{r} \xi'(r) + \xi''(r) \big)
	+ 2\beta \xi'(l) \xi'(r)  \bigg]  .
\label{Ainsphericalcoord}
\end{eqnarray}
In observation and simulations,
the ratio  $u=2$ is often taken,
 so that $\zeta(r, u, \theta)$ will have only two variables.
The 2PCF  $\xi(r)$ is involved in eq.\eqref{3PCF_02sp}.
Although $\xi(r)$ has been solved to various nonlinear orders
\cite{zhang2009nonlinear,ZhangChen2015,ZhangChenWu2019},
we shall use the observed $\xi(r)$ \cite{marin2011}
for a coherent comparison with observation.

An appropriate boundary condition is needed to solve eq.\eqref{3PCF_02sp}.
Ref.\cite{marin2011} has observed the redshift-space  $Q(s, u, \theta)$
of ``DR7-Dim" (61,899 galaxies in the  range $0.16 \leq z \leq 0.36$)
from  SDSS in the domain
$s \in [7.0 , 30.0]\, h^{-1} {\rm Mpc}$,
$\theta \in [0.1 , 3.04]$
at five respective values  $s=7,10,15,20,30\, h^{-1}$Mpc at a fixed $u=2$,
where $s$ is the redshift distance.
(See Fig. 6 and Fig. 7 of Ref.\cite{marin2011}.)
$s$ may differ from the real distance $r$ due to the peculiar velocities.
We shall this error and take $r=s$ in our computation.
From this data,
we get the fitted $Q(r, u, \theta)$,
as well as $\zeta(r, u, \theta)$ via the relation \eqref{grothpeeblesansatz},
on the boundary of the domain,
which is taken  as the  boundary  condition of eq.\eqref{3PCF_02sp}.
The effect of the delta source is absorbed
by the boundary condition \cite{ZhangLi2021,Hackbusch2017}.

We solve  eq.(\ref{3PCF_02sp})  numerically by the finite element method,
and obtain the solution $\zeta(r, u, \theta)$
and the reduced $Q(r, u, \theta)$ defined by (\ref{grothpeeblesansatz}).
To match the observational data \cite{marin2011},
using the $\chi^2$ test,
the parameters are chosen as the following:
$a^{(3)}_r \simeq -4.4\, h$Mpc$^{-1}$,
$a^{(2)}_r \simeq 0.35\, h$Mpc$^{-1}$,
$b \simeq 0.73$,
$c \simeq 0.03\, h^2$Mpc$^{-2}$,
$e \simeq -6.9$,
$Q \simeq 1.7$,
$R_a \simeq 4.1$,
$R_b \simeq -0.47$,
$k_J \simeq 0.038 \, h$Mpc$^{-1}$.
In particular,
the values of $Q$, $R_a$ and $R_b$ of the anzats
are consistent with that inferred from other surveys  \cite{Peebles1993}.
Besides, the chosen $k_J$ is also consistent
the value used  in our previously work on
 the 2PCF \cite{Zhang2007,zhang2009nonlinear,ZhangChen2015,ZhangChenWu2019}.
The  parameter $\alpha$ has not been accurately fixed
because the delta source has been absorbed into the boundary condition
 in our numeric solution \cite{ZhangLi2021,Hackbusch2017}.

\begin{figure}[htb]
\centering
\subcaptionbox{}
    {%
        \includegraphics[width = .45\linewidth]{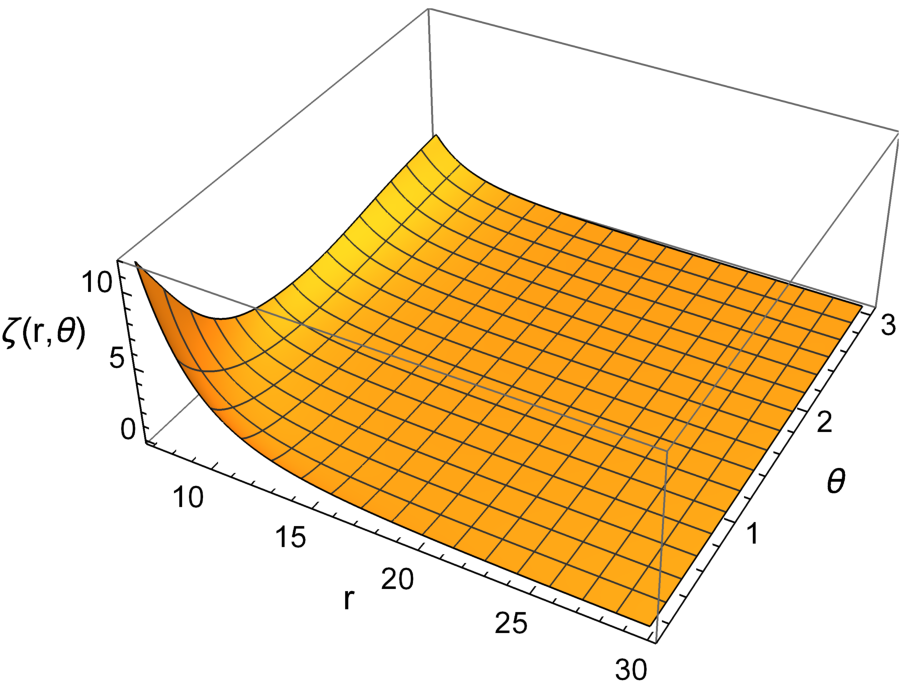}
}
\subcaptionbox{}
    {%
        \includegraphics[width = .45\linewidth]{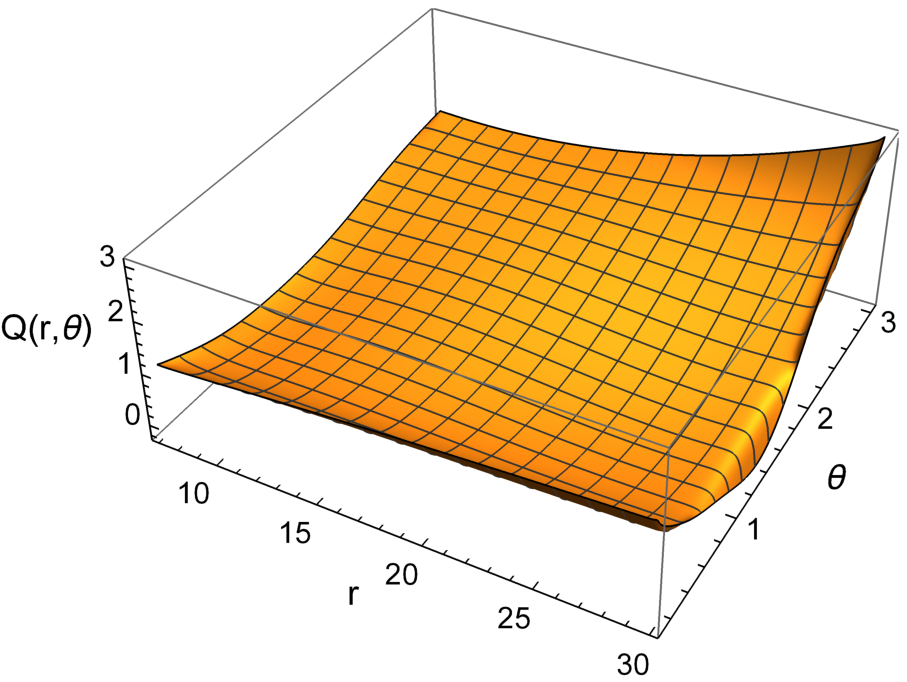}
}
\caption{
(a). The nonlinear $\zeta(r, u, \theta)$ (with fixed $u=2$)  is positive,
   has a $U$-shape along  $\theta$,
   and decreases monotonously along $r$
   up to the range $r \leq  30\, h^{-1}$Mpc of our computation.
  This behavior differs from  Fig.\ref{zetagauss} of the Gaussian solution.
(b). The reduced  $Q(r, u, \theta) \ne 1$,
exhibits a deeper $U$-shape along $\theta$,
 varies non-monotonously  along   $r$,
and its high values occur at large $r$ where $\xi(r)$ is small.
}
\label{fig3ab}
\end{figure}

Fig.\ref{fig3ab} (a)  shows the solution $\zeta(r, u, \theta)$
at fixed $u=2$  as a function of ($r,\theta$).
It is seen that $\zeta(r, u, \theta)>0$ in the range of computation,
and exhibits a shallow $U$-shape along the $\theta-$direction.
This feature is consistent with observations \cite{Guo2013,Guo2016}.
Along the $r-$direction $\zeta(r, u, \theta)$ decreases monotonously
up to  $30 h^{-1}$Mpc in the range.
The highest values of $\zeta(r, u, \theta)$ occur at small $r$,
just as   $\xi(r)$ does.
This is also expected since the correlations are stronger
at small distance due to gravity.

Fig.\ref{fig3ab} (b) shows the nonlinear reduced $Q(r, u, \theta)\ne 1$,
deviating from the Gaussianity $Q=1$.
$Q(r, u, \theta)$  exhibits a deeper $U$-shape along $\theta$,
and varies non-monotonically along $r$.
The variation along $r$ is comparatively weaker than the variation along $\theta$.
These features are consistent
with what have been observed \cite{marin2011,McBride2011a,McBride2011b}.

\begin{figure}[htbp]
\centering	\includegraphics[width=0.7\columnwidth]{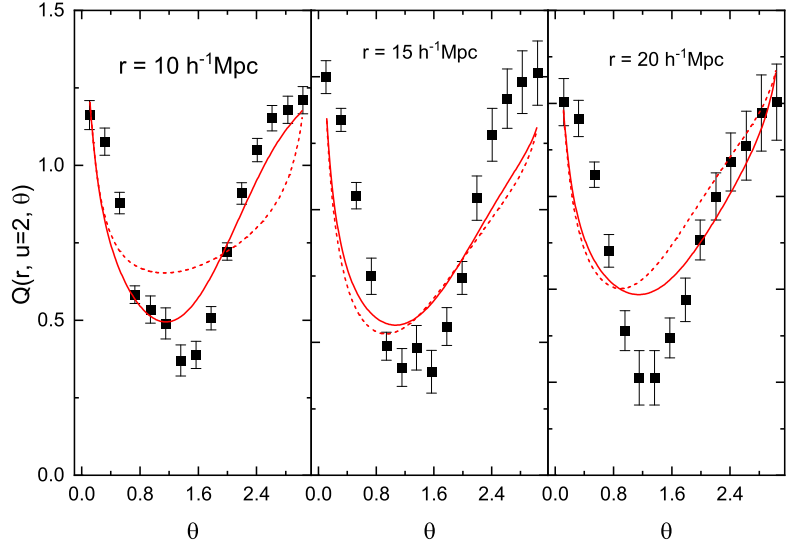}\\
	\caption{  \label{qcompare}
The solid line:   $Q(r, u, \theta)$ from eq.\eqref{3PCF_02sp}.
The dashed line: the second order solution from Fig.6 of
Ref.\cite{WuZhang2021}.
Three plots are for $r=10  h^{-1} {\rm Mpc}$,
	$15  h^{-1} {\rm Mpc}$, $20  h^{-1} {\rm Mpc}$, respectively.
The dots:  the SDSS data from
    Fig.6 and Fig.7 of Ref. \cite{marin2011},
    which are measured in the redshift space.
    Our solutions are of the real space.
    There are differences between the real and redshift spaces,
    which are neglected in this preliminary treatment here.
	}
\end{figure}
To  compare with the observational data \cite{marin2011},
Fig.\ref{qcompare} plots the solution $Q(r, u, \theta)$
as a function of $\theta$
at   $r=10, 15 ,20  \,  h^{-1} {\rm Mpc}$,  respectively.
It is seen that
$Q(r, u, \theta)$ has a $U$-shape along  $\theta=[0,3]$,
 agreeing with the data.
Overall,  the equations of 3PCF gives a reasonable account
of the data of galaxies with redshifts
 $0.16 \leq z \leq 0.36$.
For a comparison, in Fig.\ref{qcompare} we also plot the second order solution
(dashed lines).
Note that we have renormalized the parameters of
the third order solution in this paper,
the number of parameters also differs from that
of the the second order.
It is clear that the third order solution fits
the data ($\chi^2=470.9$)
 better than the second order one ($\chi^2=777.08$),
especially at small scales, and the two
solutions are close at large scales.

\section{Conclusions and discussions}

Based on the density field equation \eqref{psifieldequ},
we have derived the  equation \eqref{3PCF}
of the 3-point correlation function $G^{(3)}$ of galaxies,
up to the third order density fluctuation.
This work is a continuation of
the previous Gaussian approximation  \cite{ZhangChenWu2019},
and the second order work \cite{WuZhang2021}.

By neglecting the 5PCF,
adopting the Fry-Peebles  ansatz to deal with the 4PCF,
and the Groth-Peebles ansatz to deal with the squeezed 3PCF, respectively,
we have made  eq.\eqref{3PCF} into
the closed equation \eqref{3PCF_02}.
Aside the three parameters from the anzats,
there are six nonlinearity parameters that occur inevitably
in the perturbation treatment of a gravitating system.
We carry out renormalization of  the Jeans wavenumber and the mass.
Although the terms $(\delta\psi)^3$ are included,
nonlinear terms such as $(G^{(3)})^2$ do not appear
in eq.\eqref{3PCF_02} of $G^{(3)}$,
and higher order terms than $(\delta\psi)^3$ are needed
for $(G^{(3)})^2$ to appear.

We apply the equation to the system of galaxies,
using the boundary condition
inferred from SDSS DR7 \cite{marin2011} for a consistent comparison.
The solution $\zeta(r, u, \theta)$ exhibits a shallow $U$-shape along $\theta$,
and  decreases monotonously along $r$.
The reduced $Q(r, u, \theta)$
deviates from $1$ of the Gaussian case,
and exhibits a $U$-shape along $\theta$.
Along $r$, however,  $Q(r, u, \theta)$ varies non-monotonically,
scattering around $1$.

It is interesting that the third order solution in this paper
is quite close to the second order solution \cite{WuZhang2021},
especially at large scales.
This indicates  that
the density field theory
 with increasing orders of perturbation
provides a rather stable description of the nonlinear galaxy system.
Besides,
from the study on 3PCF and  the previous work on 2PCF,
it is seen  that the static equations of correlation functions
present a reasonable analytical account of
the galaxy distribution at small redshifts.
The future work will be application to new observational data,
and extension to the case of  expanding Universe.

\section*{Acknowledgements}

Y. Zhang is supported by NSFC Grant No. 11675165,  11633001,  11961131007,
 and in part by National Key RD Program of China (2021YFC2203100).

\end{document}